\documentclass[psfig]{article}
\input psfig.sty

\begin{document}



\title{HERMES AND THE SPIN OF THE PROTON
}

\author{H. E. Jackson \\
 Physics Division, Argonne National Laboratory, \\
Argonne, Illinois 60439,
United States
}

\maketitle

\date{}

\begin{abstract}
HERMES is a second generation experiment to study the spin structure of
the nucleon, in which measurements of the spin dependent properties of 
semi-inclusive deep-inelastic lepton scattering are emphasized. 
Data have been accumulated for semi-inclusive
pion, kaon, and proton double-spin asymmetries, as well as for high-$p_{T}$ 
hadron pairs, and single-spin azimuthal 
asymmetries for pion electroproduction and deep 
virtual Compton scattering. These results provide information on the 
flavor decomposition of the polarized quark distributions in the nucleon and
a first glimpse of the gluon polarization, while the observation of the
azimuthal asymmetries show promise for probing the tensor spin of the 
nucleon and isolating the total angular momentum carried by the quarks.  

\end{abstract}

\section{Introduction}

As Sir Isaac Newton speculated centuries ago in his treatise, {\it Optics},
``There are therefore agents in nature able to make the  particles of
bodies stick together by very strong attractions. And it is the business of
experimental philosophy to find them out.'' Indeed, one of the central 
challenges of contempory nuclear physics is achieving an understanding
of how quarks and gluons combine to make mesons, baryons, and nuclei.
One particular issue which has received much attention is how the 
constituent-parton spins combine to form the spin of the proton. This is
not a new problem. The first experiments to explore the spin structure of
the nucleon were performed 30 years ago.\cite{Hughes} 
They took the form of measurements of inclusive spin asymmetries, the basic
measurement which, until now, has provided most of our knowledge 
of nucleon spin structure.

The objective of these studies was to determine what fraction of the spin
of the nucleon was carried by the quarks. The nucleon spin can be decomposed
conceptually, into the spin of its constituents according to 
the equation
\begin{equation}
\langle s{^N}{_z}\rangle = \frac{1}{2}  =\frac{1}{2}{\Delta\Sigma
}+{\Delta G}
+{\Delta L_{z}},
\label{nuclspin}
\end{equation}
where the three terms are the quark and gluon spins, and the total orbital
angular momenta of the quarks and gluons, respectively.
The original expectation, based on the constituent quark model was that the
intrinsic spin of the valence quarks provided the total spin, 
ie $\Delta\Sigma =1$. 
More realistic calculations with current quarks\cite{Jaffe} resulted in
a ``canonical value'' of $\Delta\Sigma\approx 2/3$. 
Then came the first experiments
with polarized deep-inelastic lepton scattering at SLAC\cite{Hughes} 
and then CERN\cite{CERN} which
led to the conclusion that $\Delta\Sigma \approx 0.1-0.2$. The prospect that
the fraction of the nucleon spin carried by the quarks was so small
provoked what was called the ``spin crisis''.

With these indications of the complexity of the spin structure, it was 
quickly realized that a simple leading order(LO) 
analysis which did not include 
contributions from gluons 
was naive. More recent next-to-leading order(NLO)
treatments provide a picture more consonant with our present understanding
of QCD. The focus has been on the spin dependent structure function for the
proton, $g_{1}(x,Q^{2})$, given by\cite{altarelli}
\begin{equation}
g_{1}(x,Q^{2})=\frac{\langle e^{2}\rangle}{2}[C_{NS}\otimes\Delta q_{NS}+
C_{S}\otimes\Delta\Sigma+2n_{f}C_{g}\otimes\Delta g]
\label{g1}
\end{equation}
where $\langle e^{2}\rangle =n_{f}^{-1}\Sigma_{i=1}^{n^{f}}e^{2}_i$, 
$\otimes$ denotes convolution
over $x$, and $\Delta q_{NS}$ and $\Delta\Sigma$ are the nonsinglet and 
singlet quark distributions, and $\Delta g$ is the polarized gluon 
distribution, respectively. 
Here $x$ is the usual Bjorken scaling variable, $Q^2$ is the squared
four momentum transfer, and $n_f$ is the number of quark flavors. 
The coefficient functions,
$C_{NS}$, $C_S$, and $C_g$, have
been computed up to next-to-leading order\cite{Kodaira} in $\alpha_{s}$.
At NLO they as well as their associated parton distributions depend on 
the renormalization and factorization schemes. While the physical
observables are scheme independent, parton distributions will be 
maximally scheme dependent, but related from scheme to scheme by well-defined 
relationships. In a recent NLO analysis\cite{adeva} of available data 
for $g_{1}$, the SMC group presented results for the Adler-Bardeen(AB) 
scheme, in
which case the first moment of $g_{1}$,
$\Gamma_{1}(Q^{2})$, is given by
\begin{equation}
\int_{0}^{1} dxg_{1}(x,Q^{2})=\frac{\langle e^2\rangle}{2}
[C_{NS}(1,\alpha_{s}(t))
{\Delta}q_{NS}(1)+C_{s}(1,\alpha_{s}(t))a_0(Q^{2})]
\label{1stmom}
\end{equation} 
where we have followed the notation of Ref.\cite{altarelli} in denoting
moments of coefficient functions and parton densities as 
$f(N)=\int_{0}^{1}dx\,x^{N-1}f(x)$. 
In the same scheme the singlet axial charge, $a_{0}$, is
\begin{equation}
a_{0}= \Delta\Sigma (1,Q^{2})-3\frac{\alpha _{s}
    (Q^{2})}{2\pi}\Delta g(1,Q^{2}).
\label{a0}
\end{equation}  
The SMC group finds that the analysis of the $Q^2$ 
evolution of the world data base
gives an singlet axial charge, $a_{0}=0.23\pm 0.07(stat)\pm 0.19(stat)$, 
and a gluonic first moment, $\Delta g(1,1GeV^{2})=0.99^{+1.17}_{-0.31}
(stat)^{+0.42}_{-0.22}(syst)^{+1.43}_{-0.45}(th)$. 
The resulting value of the 
singlet quark distribution is $\Delta\Sigma = 0.38$. However, the
large uncertainties in $\Delta g$ preclude a precise determination of
$\Delta\Sigma$ in the absence of direct measurements of $\Delta g$
\cite{adams}.  An additional issue in the analysis of the inclusive
data from these experiments, is their sensitivity to SU(3) symmetry
breaking. In a recent study, Leader and coworkers\cite{leader} find
that the strange quark and gluon polarizations vary rapidly with
variations in the SU(3) octet axial charge. Clearly, these flavor separated
polarizations are strongly dependent on the assumption of SU(3)
symmetry.  

The key to further progress is more specific probes of the 
individual contributions of Eq.~(\ref{nuclspin}) to the proton spin. 
Determination of the 
polarization of the gluons is clearly of very high priority, and in addition, 
a more precise measurement
will eliminate a major current ambiguity in the implications
of existing inclusive data. A more 
direct determination of the strange quark polarization, $\Delta s$, will avoid 
the need for the use of data from hyperon decay and the assumption of 
SU(3) symmetry. Measurements which are sensitive to quark charges will
allow the separation of quark and antiquark polarizations. The HERMES 
experiment attempts to achieve these objectives, by emphasizing 
semi-inclusive DIS in which a $\pi ,K,$ or $p$ is observed in
coincidence with the scattered lepton. The added dimension of flavor in
the final hadron provides a valuable probe of the  flavor dependence 
and other features of polarized parton distributions. In the sections which 
follow, we present a brief description of the HERMES experiment followed 
by  reports on recent results on the flavor decomposition of polarized
parton distributions, gluon polarization, transverse spin physics, and
deep-virtual Compton scattering. Indeed, such studies of semi-inclusive 
deep-inelastic scattering appear to mark a major advance in unraveling
the spin structure of the proton. 

\section{The HERMES Experiment}

Deep inelastic scattering events are generated in the HERMES experiment
by the interaction of the polarized lepton beam of the 
HERA accelerator at DESY
with polarized target gases which are injected into a 40 cm long, tubular
open-ended storage cell\cite{stewart} located 
at an interaction point on the lepton 
orbit. The lepton beam is self polarized by the Sokolov-Ternov 
effect\cite{sokolov} with
a polarization time which is typically about 20 minutes. The beam 
polarization is measured continuously with Compton backscattering of 
circularly polarized laser beams\cite{barber,beckmann}. 
The beam polarization is routinely about
0.55. Spin rotators in the ring provide longitudinal polarization at the
interaction point. The polarized target gases, atomic H or D
are generated by an atomic-beam source based on Stern-Gerlach separation
which provides an areal density of about $2\times 10^{14}$ 
atoms/cm$^{2}$ for H and 
$7\times 10^{13}$ atoms/cm$^{2}$ for D. The nuclear polarization is measured
with a Breit-Rabi polarimeter\cite{baum}
 and the atomic fraction with a target gas 
analyser\cite{simani}. The target polarization is 
reversed within short time intervals
to minimize systematic effects. The relative lumimosity is 
measured\cite{benisch} by 
detecting Bhabha- or Mott-scattered target electrons in coincidence
with the scattered lepton, in a pair of NaBi(WO$_{4}$)$_{2}$ electromagnetic
calorimeters.

\begin{figure}
\centerline{\psfig{file=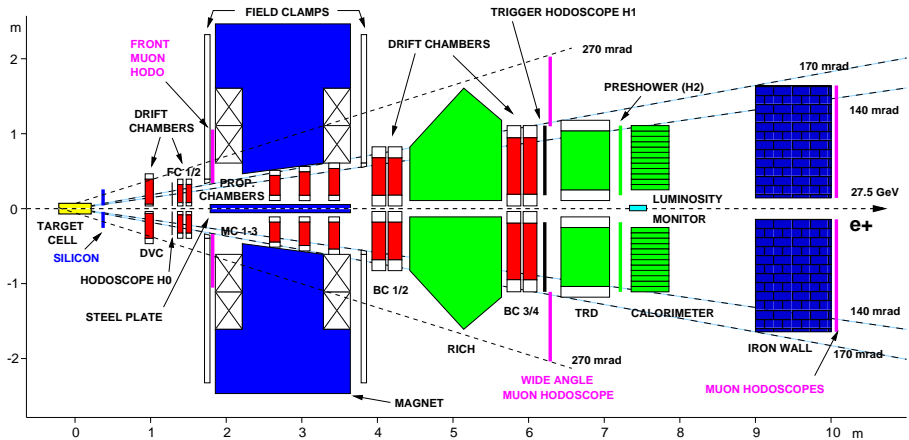,width=12cm}}
\vspace*{8pt}
\caption{The HERMES Spectrometer.}
\label{spectr}
\end{figure}

The HERMES spectrometer\cite{spec},
shown in Fig.~(\ref{spectr}) is a forward angle open geometry system
consisting of two halves which are symmetric about a central horizontal
shielding plate in the spectrometer magnet. 
A fly's eye calorimeter and a transition radiation detector furnish 
clean separation of hadrons and leptons. Identification of $\pi's,K's$ and 
$p's$ is accomplished by means of a novel dual-radiator ring-imaging
Cerenkov counter(RICH)\cite{rich} which 
is located between the rear tracking chambers. The combination of radiators
consisting of a wall of clear aerogel and a gas volume of $C_{4}F_{10}$ 
provide clean particle identification over almost the full acceptance of
HERMES, i.e. 2-15 GeV. The scattered leptons and hadrons 
produced within an angular acceptance of $\pm$ 170 mr horizontally,
and 40 - 140 mrad vertically are detected and identified. Typical kinematics
for studies of DIS are $E=27.5$ GeV for the incident lepton, 
$x>0.02$ where $x=Q^{2}/2M\nu$ is the Bjorken scaling 
variable, $0.1$ GeV$^{2}$ $<Q^{2}
<15$ GeV$^{2}$ with $-Q^{2}$ the square of the momentum transfer, and 
$\nu < 24$ GeV where $\nu =E-E'$ with $E(E')$ is the energy 
of the incoming(scattered) lepton in the target rest frame. 
To insure that hadrons 
detected are in the current fragmentation region, cuts of $z=E_{h}/\nu >0.2$
and $x_{F}\approx 2p_{parallel}^{h}/W > 0.1$ are imposed, where
$W=\sqrt{2M\nu +M^{2}-Q^{2}}$ is the invariant mass of the photon-proton
sytem.

\section{Flavor Decomposition}

One of the principal goals from the earliest days of the HERMES experiment
has been to use  semi-inclusive deep-inelastic scattering 
to determine the separate contributions 
$\Delta q_{f}(x)$ of the quarks and antiquarks of flavor $f$ 
to the total spin of the nucleon. By means of the technique of flavor 
tagging, individual spin contributions can be determined directly from spin
asymmetries of hadrons with the appropriate flavor content. For example, 
the spin asymmetry of $K^{-}$, an all sea object, will have a high 
sensitivity to the polarization of the quark sea. The measured semi-inclusive
spin asymmetry, $A_{\parallel}^{h}$, and the corresponding photon-nucleon
asymmetry, $A_{1}^{h}$, for the hadron of type $h$ are given by
\begin{equation}
A^{(h)}_{\parallel}=\frac{N_{(h)}^{\uparrow\downarrow} - N_{(h)
}^{\uparrow\uparrow}}{N_{(h)}^{\uparrow\downarrow} +
N_{(h)}^{\uparrow\uparrow}} , \ \ A^{h}_{1}=\frac{A^{h}_{\parallel}}
    {D(1+\eta\gamma )},
\label{asym}
\end{equation}
where, for simplicity, we assume unity beam and target polarizations and
constant luminosity. Here $D$ is the depolarization factor for the virtual
photon, $\eta$ is a kinematic factor\cite{aira1}, and
$N^{\uparrow\uparrow}$$(N^{\uparrow\downarrow})$ are the number of DIS events
with coincident hadrons for target polarization parallel (anti-parallel) to
the beam polarization. In leading order QCD assuming the validity of
factorization, one can write the semi-inclusive DIS cross section, 
$\sigma^{h}(x,Q^{2},z)$, to produce a hadron with energy fraction, 
$z=E_{h}/\nu$ as
\begin{equation}
\sigma^{h}(x,Q^{2},z)\propto \Sigma_{f}e^{2}_{f}q_{f}(x,Q^{2},z)
D_{f}^{h}(x,Q^{2})
\label{sigmah}
\end{equation} 
where the sum is over quark and antiquark types $f=(u,\overline{u},
d,\overline{d},s,\overline{s})$. $E_{h}$ is the energy of the hadron.
The quark charge, $e_{f}$, is in units 
of the elementary charge. In this approximation, 
\begin{eqnarray}
A^{h}_{1}(x,z) & = & \frac{\int_{z_{min}}^{1}dz\sum_{f}e^{2}_{f}
  {q_{f}(x)}\cdot{D^{h}_{f}(z)}}
  {\int_{z_{min}}^{1}dz\sum_{f'}e^{2}_{f'}
  {q_{f'}(x)}\cdot{D^{h}_{f'}(z)}}
  \cdot{\frac{\Delta q_{f}(x)}{q_{f}(x)}} 
  \cdot{\frac{1+R(x,Q^{2})}{1+\gamma^{2}}} \\
   & = & \sum_{f}{P^{h}_{f}(x,z)}
  \frac{\Delta q_{f}(x)}{q_{f}(x)}\cdot{\frac{1+R(x,Q^{2})}{1+\gamma^{2}}}.
\label{purity}
\end{eqnarray}
The quantities, ${P^{h}_{f}(x,z)}$, are the integrated 
purities\cite{funk,nicz}
which are defined by Eq.~(\ref{purity}). They are 
\begin{figure}
\centerline{\psfig{file=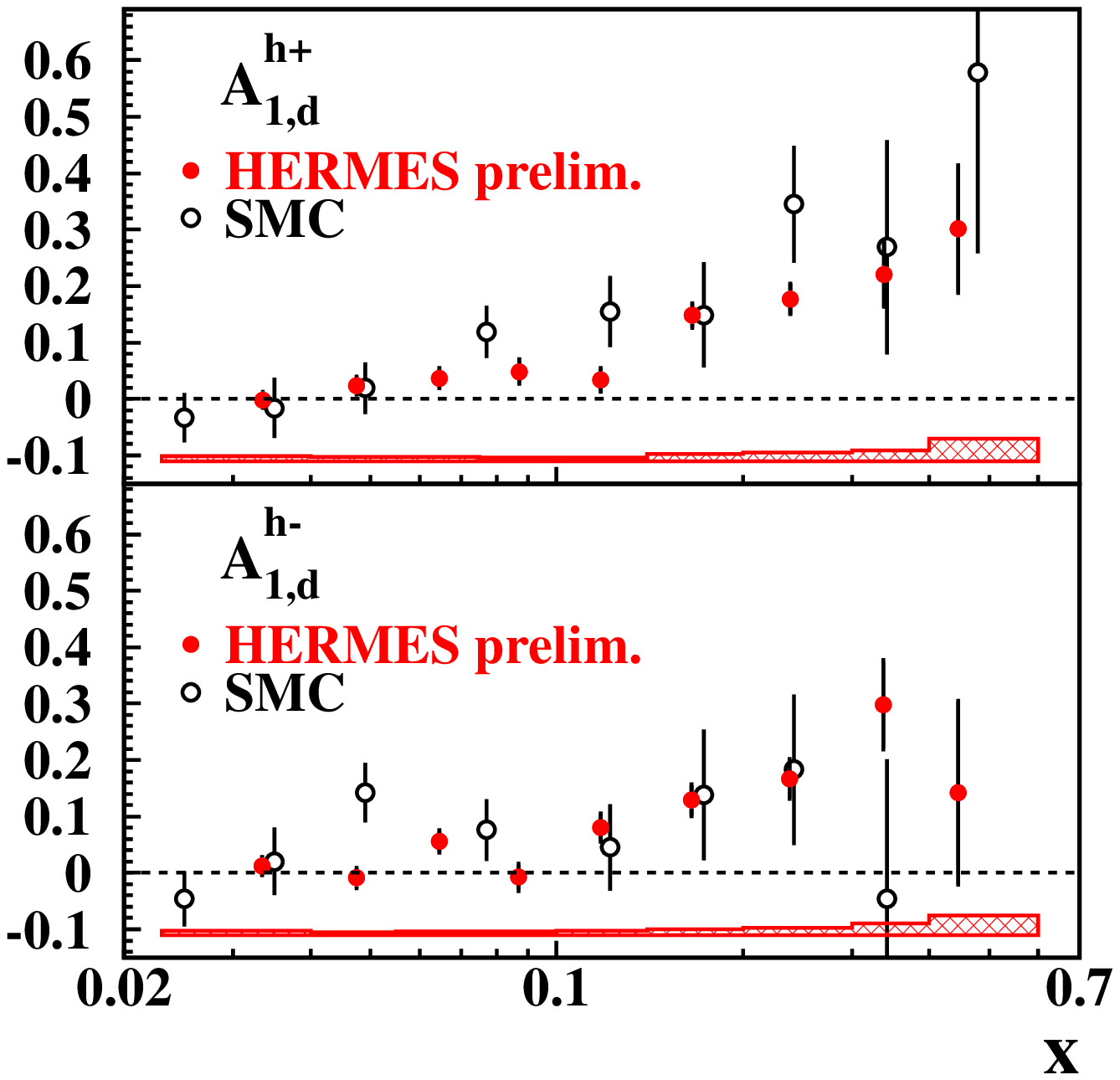,width=4cm}
\psfig{file=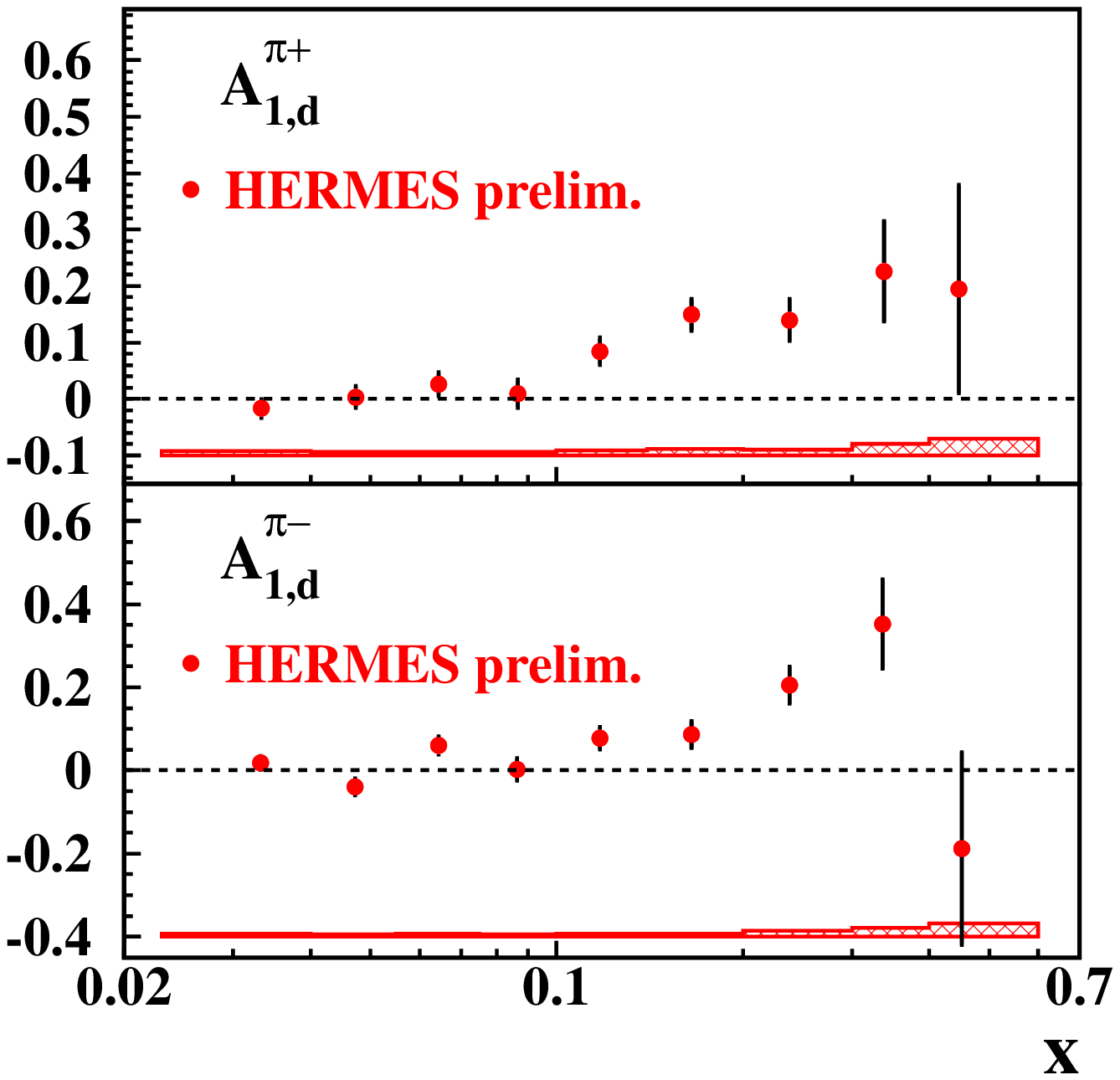,width=4cm}
\psfig{file=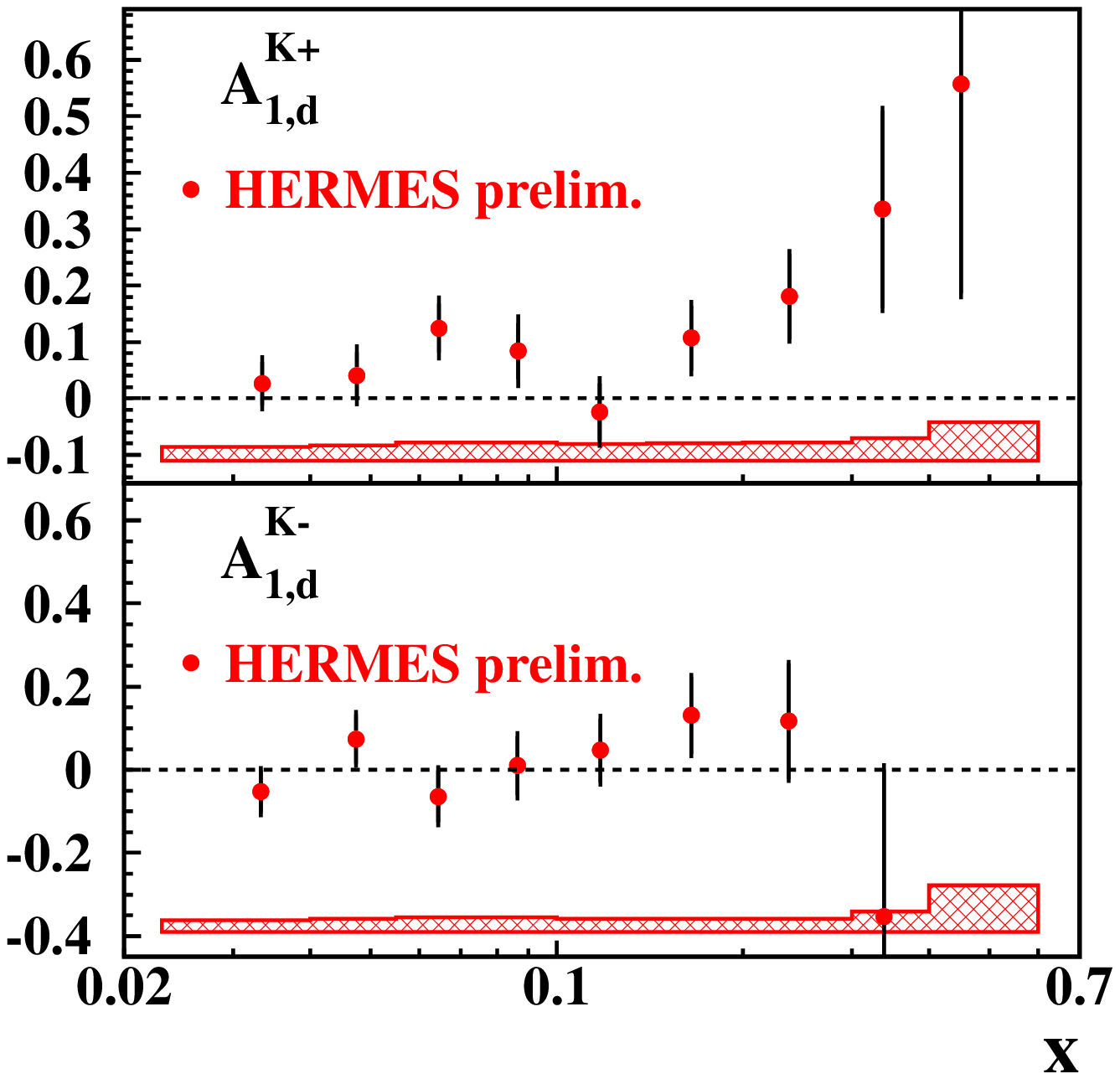,width=4cm}}
\vspace*{8pt}
\caption{Semi-inclusive hadron, pion and kaon asymmetries for a
deuterium target. The hadron asymmetries are compared with data from
the SMC collaboration. The error bars of the HERMES data are statistical
and the bands are systematic uncertainties.
These data form the deuteron portion
of the data base 
used in the purity analysis described in the text.}
\label{a1ds}
\end{figure}
spin-independent quantities in leading order
and represent the probability that the quark, $q_{f}$
was struck in the DIS event. The ratio $R=\sigma_{L}/\sigma_{T}$ 
of the longitudinal to transverse photon cross section corrects for 
the longitudinal component that is included in experimental parameterizations
of $q_{f}(x,Q^{2})$ but not in $\Delta q_{f}(x,Q^{2})$. The term $\gamma=
\sqrt{Q^{2}}/\nu$ is a kinematic factor. By incorporating the correction
factor in the purities, one can rewrite Eq. ~(\ref{purity}) in a matrix form
as
\begin{equation}
A(x)=P(x)\cdot Q(x)
\label{matrix}
\end{equation}
where $A(x)$ becomes a vector whose elements are 
all the integrated measured asymmetries
which are to be included in the analysis. The $Q(x)$ vector contains the
quark and antiquark polarizations. These quantities are now connected by the
purity matrix which contains the effective integrated purities. The 
determination of the quark polarizations from the experimentally measured
spin asymmetries is reduced to the task\cite{decomp}
of inversion of Eq.~(\ref{matrix}) to obtain $Q(x)$.
\begin{figure}
\centerline{\psfig{file=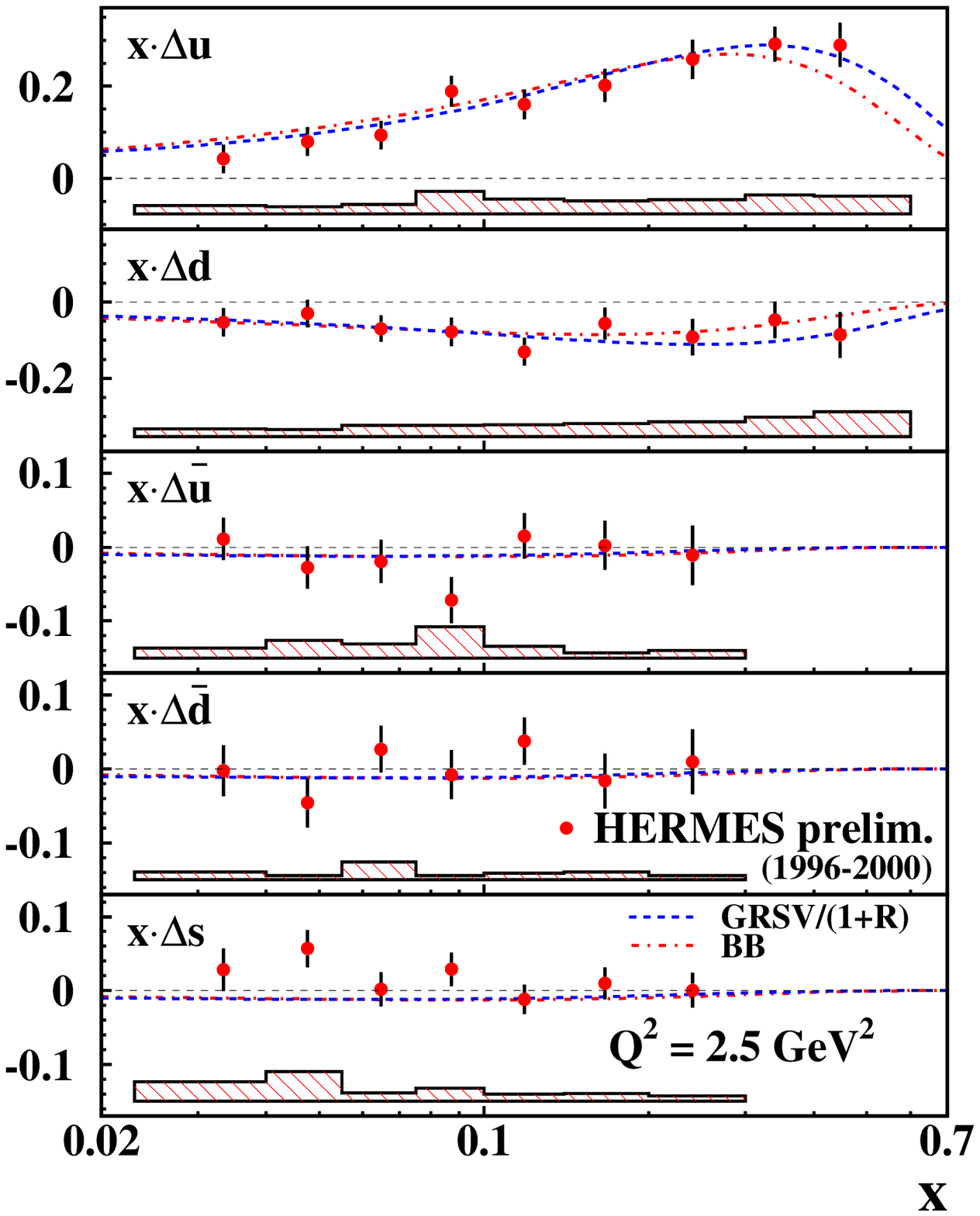,width=10cm}}
\vspace*{8pt}
\caption{The $x$-weighted polarized quark densities. The plots show
a five parameter fit to the data assuming a symmetric strange sea 
polarization. The data have been evolved to a common $Q^{2}=2.5$ GeV$^2$.
The dashed line shows a GRSV parameterization, and the dashed-dotted
curve an alternate parameterization of Bluemlein and Boettcher 
(hep-ph/0203155).}
\label{flav}
\end{figure}

The purity formalism has been used in the HERMES analysis to make a 
flavor decomposition into polarized quark distributions for 
$u,\overline{u},d,\overline{d}$, and $s+\overline{s}$.
For the first time, a global analysis of
inclusive spin asymmetries and semi-inclusive spin asymmetries for 
$\pi^{+},\pi^{-},K^{+},$ and $K^{-}$ 
has been carried out for longitudinally polarized targets of
hydrogen, and deuterium. The measured spin asymmetries $A_{1}^{h}
(x,Q^{2},z)$ were integrated in each $x$ bin over the corresponding
$Q^{2}$-range and the $z$-range from 0.2 to 1 to yield $A_{1}^{h}(x)$. 
The data for $A_{1}^{h}(x)$ obtained with the deuterium
target are shown in Fig.~(\ref{a1ds}). There the semi-inclusive data are 
compared to earlier results from the SMC collaboration\cite{smci}.
The purities were obtained with a Monte Carlo calculation
which used CTEQ5 leading order parton distributions\cite{cteq5} and a 
LUND fragmentation model\cite{lund} tuned to HERMES kinematics. 
The simulation included effects of the acceptance
of the experiment. Systematic uncertainties were estimated by varying
the fragmentation parameters and using alternative 
parton distributions\cite{grv98}. 
\begin{figure}[b!]
\centerline{\psfig{file=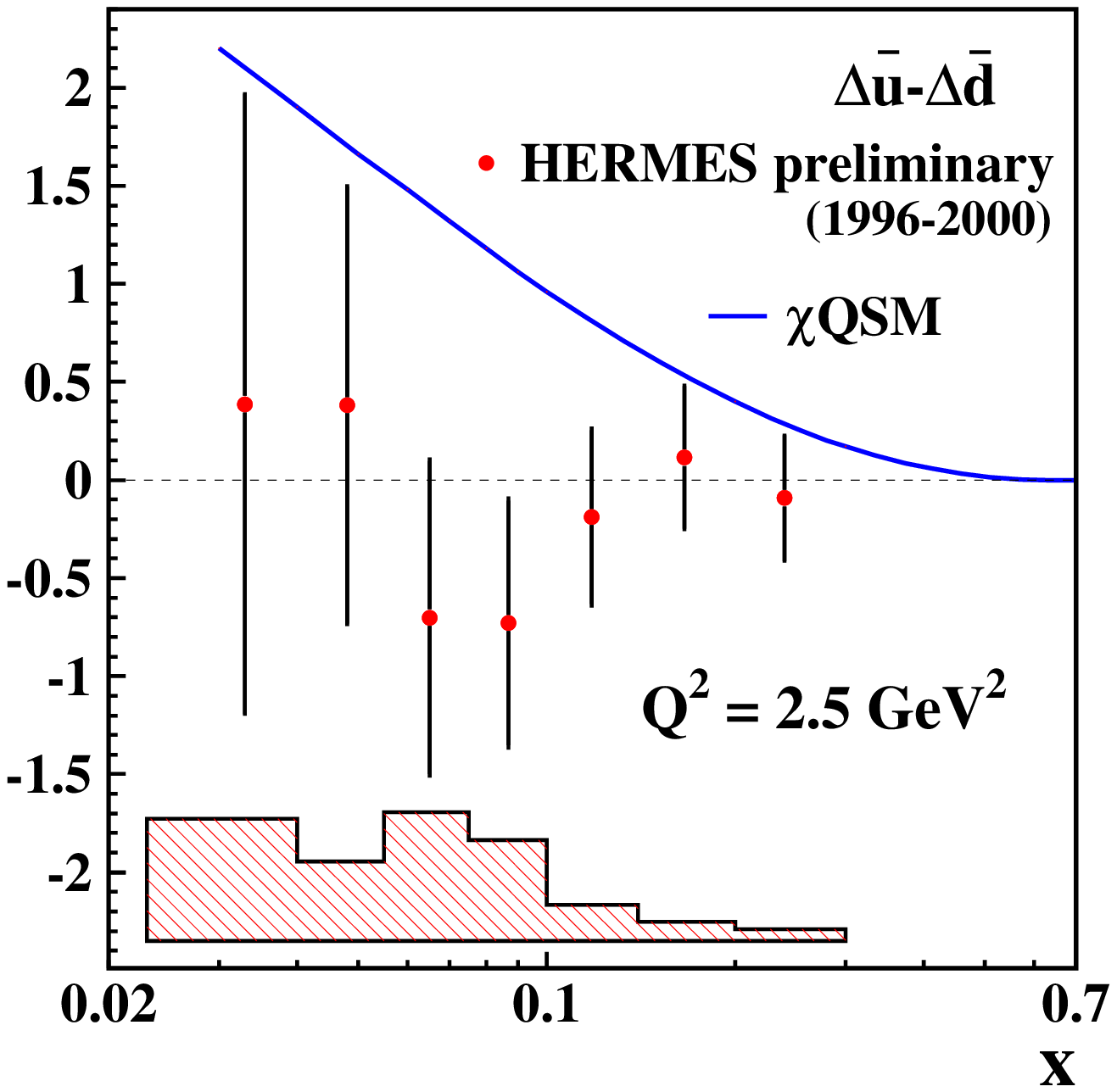,width=8cm}}
\vspace*{8pt}
\caption{Flavor asymmetry $\Delta u-\Delta d$ of the light sea extracted
from the HERMES five-component purity analysis. The curve describes
a prediction of the $\chi$QSM model (see text). The error bars give
statistical uncertainties and the shaded band the systematic error.}
\label{chib}
\end{figure}

The results of the decomposition obtained by solving Eq.~(\ref{matrix})
are presented in in Fig.~(\ref{flav}). A symmetric
strange sea polarization was assumed, i. e.
$\Delta s/s=\Delta \overline{s}/\overline{s}$. The general features of
quark densities follow those of earlier decompositions\cite{aker2,adev2}.
The $u$-quarks show a strong positive polarization, while the $d$-quarks
have a substantial negative polarization. The non-strange sea quarks are
not significantly polarized. However, the strange sea appears to be 
positively polarized, contrary to the conclusions drawn within leading
order QCD\cite{fili} from earlier inclusive data. The triplet strength
$\Delta q_{3}=\Delta u-\Delta d$ extracted from the HERMES data is in 
agreement with the Bjorken sum rule. The polarization of the strange sea 
can be extracted directly from the same data set by means of a purity
analysis which uses only two spin asymmetries, $A_{1}^{D}(x)$ and
$A_{1,D}^{K^{+}+K^{-}}(x)$. For this analysis, fragmentation functions 
from $e^{+}e^{-}$ collider experiments can be used to calculate purities.
This method measures the quantity $\Delta s+\Delta\overline{s}$ with no
assumption about strange sea symmetry, and provides an independent check of
the result from the five-component decomposition. The results obtained show
the same trend of positive strange sea polarization. This unexpected
result poses a challenge to our understanding of the quark sea. To the
extent that the sea arises from gluon splitting, one would expect
from the result for the strange sea that
$\Delta\overline{u}\approx\Delta\overline{d}>0$. 

The data do not support recent conjectures of a strong breaking of the
flavor symmetry of the light sea. The results for the quantity
$\Delta u-\Delta d$ are shown in Fig.~(\ref{chib}) together with a 
prediction based on the chiral quark soliton model\cite{dres}
($\chi$QSM). Although
the statistics are limited, the data indicate that any flavor asymmetry
in the nonstrange sea is substantially smaller that the prediction of
the $\chi$QSM. In addition, the conclusion that a large negative polarization
of the strange sea is the explanation for the violation of the
Ellis-Jaffe sum rule in inclusive DIS is ruled out by the new results 
from HERMES. Analysis of the HERMES data is continuing. A point of
major interest which will emerge is a determination of the octet 
strength $\Delta q_{8}=\Delta u+\Delta d-2\Delta s$ as a test of
SU(3) symmetry. The HERMES results represent the first complete flavor
decomposition of the quark contribution in Eq.~(\ref{nuclspin}) to the spin of
the nucleon.

\section{Gluon Polarization}

\begin{figure}
\centerline{\psfig{file=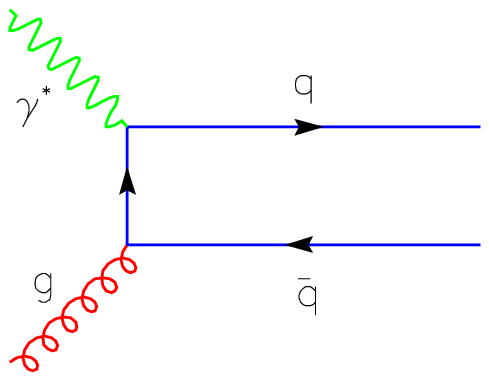,width=4cm}\hspace*{2cm}
\psfig{file=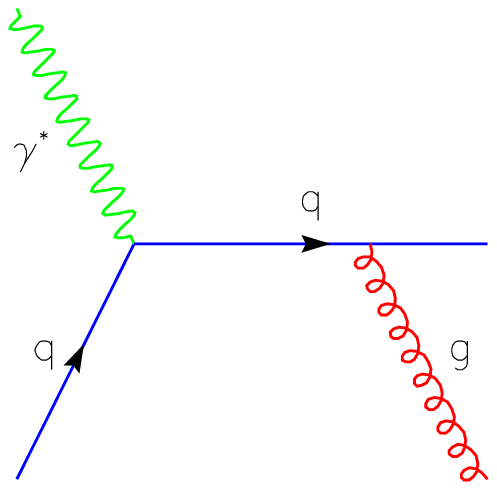,width=3.3cm}}
\vspace*{8pt}
\caption{Feyman diagrams for processes which are major contributors
to production of high-$p_T$ pairs, photon-gluon fusion(left) and 
QCD Compton scattering(right).}
\label{fusion}
\end{figure}
Indications from the analysis of the $Q^{2}$ evolution of inclusive spin 
asymmetries\cite{altarelli,adeva} suggest that the first moment of the
gluon polarization, $\Delta g(1,Q^{2})$ is positive and large, although
the precise value is poorly constrained. A direct measurement of
$\Delta g(1,Q^{2})$ is of high priority in understanding the proton spin, 
and much effort is focused on its measurement. A number of processes 
have been suggested as probes of this quantity. The photon gluon fusion
process, shown in Fig.~(\ref{fusion}) is
an obvious choice. Two experimental signatures of this process are charm
production and dijet production with high transverse momentum, $p_T$.
Both have been used to measure directly the unpolarized gluon structure
function, $g(x_{g},Q^{2})$. Bravar {\it et al.}\cite{bravar} have noted that
at the lower energies characteristic of fixed target experiments, high-$p_T$
hadrons can serve as {\it pseudo} jets in a measurement of $\Delta g(1,Q^{2})$.
The spin asymmetry asymmetry associated with this process is expected
to provide a large sensitivity to $\Delta g(x_{g},Q^{2})$.

\begin{figure}
\centerline{\psfig{file=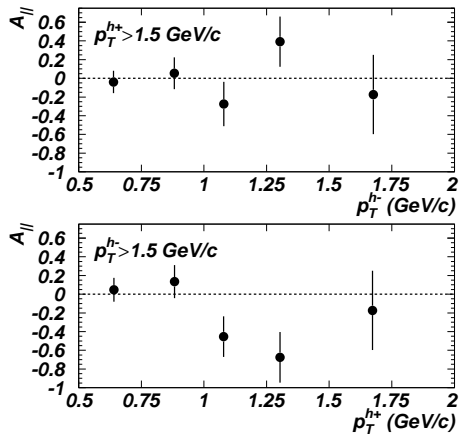,width=6
cm}}
\vspace*{8pt}
\caption{$A_{\parallel}(p_{T}^{h^{+}},p_{T}^{h^{-}})$ for $p_{T}^{h^{+}}
>1.5$ GeV/c (top) and for $p_{T}^{h^{-}}>1.5$ GeV (bottom). The error bars
represent the statistical uncertainty.}
\label{aparallel}
\end{figure}

HERMES has exploited this sensitivity by measuring the spin asymmetry in 
photoproduction of pairs of high-$p_T$ hadrons produced on a polarized
hydrogen target. Events are selected by
requiring at least two hadrons of opposite charge with an invariant mass
assuming both hadrons to be pions of $M(2\pi )> 1.0$  GeV/c$^2$ to suppress
contributions from vector mesons. The observation of the scattered lepton
is not required in the event, thus allowing inclusion of the near real 
photoproduction region($Q^{2}\approx 0$) which dominates the measured
cross section. Fig.~(\ref{aparallel}) presents $A_{\parallel}$ as measured
for the highest $p_T$ values accessible in HERMES. Here events are selected if
they contain hadron pairs of opposite charge, each with $p>4.5$ GeV/c, and 
a transverse momentum $p_{T}>0.5$ GeV/c 
where $p_T$ is defined as the momentum
transverse to the incident beam. In the top(bottom) panel the 
positive(negative) hadron was 
required to have a $p_{T}>1.5$ GeV/c and $A_\parallel$ 
is plotted as a function of the $p_T$ of the opposite charge. For the case
where the negative hadron has $p_{T}>1.5$ GeV/c, a substantial negative 
asymmetry is observed when $p_{T}>1.0$ GeV/c for 
the positive hadron. This negative 
asymmetry is to be contrasted with the positive asymmetries expected from
DIS on protons, or the small positive asymmetries associated with diffractive
production of vector mesons. Combining the data of the two panels of
Fig.~(\ref{aparallel}) over the the bins where $p_{T}^{h_{1}}>
1.5$ GeV/c and $p_{T}^{h_{2}}>1.0$ GeV/c where $h_1$ signifies the hadron
with the higher $p_T$ yields a negative asymmetry $A_{\parallel}=-0.28
\pm 0.12(stat.)\pm 0.02(syst)$. When the requirement $p_{T}^{h_{1}}>
1.5$ GeV/c is relaxed, $A_{\parallel}$ is consistent with zero.    

While a number of processes can, in principle, contribute the asymmetry
for high-$p_T$ hadron pairs, Monte-Carlo simulations using LEPTO and PYTHIA 
event generators\cite{airapetian} show that only photon-gluon fusion (PGF)
and the QCD Compton effect (QCDC) generate significant asymmetries 
at the kinematics of Fig.~(\ref{aparallel}). In this
case the experimental asymmetry takes the form
\begin{equation}
A_{\parallel}=(A_{PGF}f_{PGF}+A_{QCDC}f_{QCDC})D,
\label{apar}
\end{equation}
where $f_i$ is the unpolarized fraction of events from process $i$ and 
$D$ is the virtual photon depolarization parameter.  
In the HERMES analysis, the asymmetries $A_{i}'s$ have been approximated
by products of hard-process asymmetries and parton polarizations. These
hard-process asymmetries $\hat{a}_{PDF}=\hat{a}(\gamma g\rightarrow q
\overline{q})$ and $\hat{a}_{QCDC}=\hat{a}(\gamma g\rightarrow qg)$ are 
calculable in (LO) QCD\cite{font}. The results are $-1$ and $\approx 0.5$
respectively for the kinematics of this measurement and independent of
flavor. The measured asymmetry can then be expressed in terms of the 
total quark polarization $\Delta q/q$ and the gluon polarization 
$\Delta G/G$ as
\begin{equation}
A_{\parallel}\approx \left(\hat{a}_{PGF}\frac{\Delta G}{G}f_{PGF}+
\hat{a}_{QCDC}\frac{\Delta q}{q}f_{QCDC}\right) D,
\label{apar2}
\end{equation}
\begin{figure}
\centerline{\psfig{file=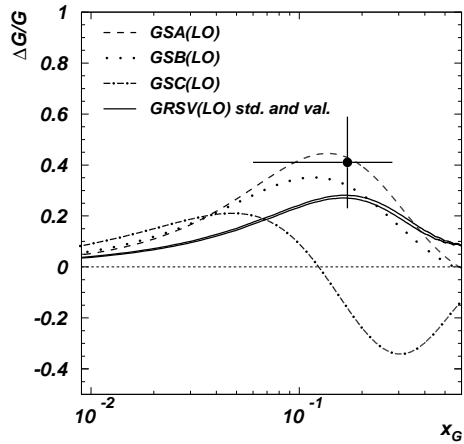,width=6cm}}
\vspace*{8pt}
\caption{The extracted value of $\Delta G/G$ compared with
phenomenological QCD fits to a subset of the world's data on
$g_{1}^{p,n}(x,Q^{2})$. 
The curves are from 
Ref's. 33,34
evaluated at a scale of $2$ (GeV/c)$^{2}$. The error indicated on 
$\Delta G/G$ represents statistical and experimental systematic
uncertainties only; no theoretical uncertainty is included.}
\label{fit1}
\end{figure}
where the kinematic dependences have been suppressed. Eq.~(\ref{apar2})
was used to extract values of $\Delta G/G$ for each of the four values
of $A_\parallel$ measured at $p_{T}^{h^2}>0.8$ GeV/c which are
obtained by averaging the corresponding asymmetries shown in 
Fig.~(\ref{aparallel}). The resulting values were then averaged to obtain
the value $\Delta G/G = 0.41\pm 0.18(stat)\pm 0.03(syst)$ for
$\langle x_{G}\rangle =0.17$ and $\langle\hat{p}_{T}^{2}\rangle =
2.1$ (GeV/c)$^2$. This value of $\Delta G/G$ 
is compared with a number of
phenomenological LO QCD fits to world data on 
$g_{1}(x,Q^{2})$\cite{gehr,gluck} in Fig.~(\ref{fit1}). 
The horizontal error bar represents
the standard deviation of the $x_G$ distribution which corresponds to the
kinematic range for the hadrons measured, as given in the Monte Carlo 
simulation. Measurements have recently been extended to the study of
high-$p_T$ kaon pairs for which the sensitivity to $\Delta G/G$ is 
enhanced by the suppression of fragmentation of non-strange quarks into 
strange hadrons. As was the case for unidentified hadron pairs, for kaon
pairs negative spin asymmetries are observed when both kaons have 
$p_T > 1.0$ GeV/c. Again the data suggest a substantial positive gluon
polarization.

\section{Transverse Spin Physics}

Three structure functions are required to provide
a complete description of the quark structure of the proton at leading
order. They are the unpolarized structure function, $f_{1}(x,Q^{2})$, the
longitudinal spin structure function, $g_{1}(x,Q^{2})$, and a transverse
structure function (transversity), $h_{1}(x,Q^{2})$ which measures the
quark spin distribution perpendicular to its momentum at infinite momentum.
Because $h_{1}(x,Q^{2})$ is chiral odd, it is not measurable in inclusive
DIS. However, it is of considerable intrinsic interest because of its 
unique properties. At low scales, $Q^{2}\approx 1$ GeV$^2$, most theories give
$h_{1}(x,Q^{2})\approx g_{1}(x,Q^{2})$,
while the first moment of $h_{1}(x,Q^{2})$,
the ``tensor charge'' includes only valence contributions. The $Q^{2}$
evolution of $h_{1}(x,Q^{2})$ is much simplier than that of its LO brothers,
because it does not couple to gluons. It is very much a valence quantity.
Transversity can be probed\cite{coll} by measuring the azimuthal
distribution of hadrons produced in polarized DIS, and with highest 
sensitivity from a  transversely polarized target.
\begin{figure}
\centerline{\psfig{file=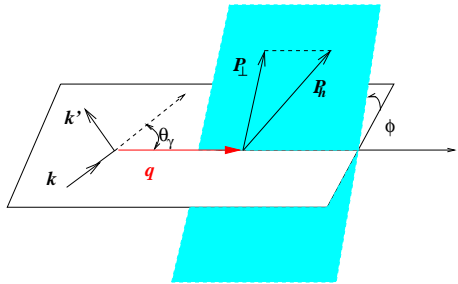,width=6cm}}
\vspace*{8pt}
\caption{Kinematics for pion electroproduction in semi-inclusive DIS.}
\label{azim1}
\end{figure}

In spite of its odd chirality, $h_{1}(x,Q^{2})$ is accessible in 
semi-inclusive DIS if combined
with a fragmentation process which also has a chiral
odd structure. This chiral odd structure results from the Collins 
effect\cite{art}, i.e. a correlation between the axis of transverse spin
and $\vec{p}_{\pi}\times\vec{q}$, a chiral odd correlation. Single spin
asymmetries are well known in proton-induced pion production from the data
of FNAL experiment E704\cite{e704} at 200 GeV. More recent theoretical
calculations by Mulders and Tangerman\cite{muld} have established, at
LO, that sin($\phi$) and sin($2\phi$) azimuthal variations in pion
single spin asymmetries signal chiral-odd fragmentation. This establishes
the very exciting possibility that spin asymmetries such as $A^{sin(\phi)}
_{p\uparrow}(p_{t})>0$ can be used as quark polarimeters for measurements
of transversity in the proton.     
 
\begin{figure}
\centerline{\psfig{file=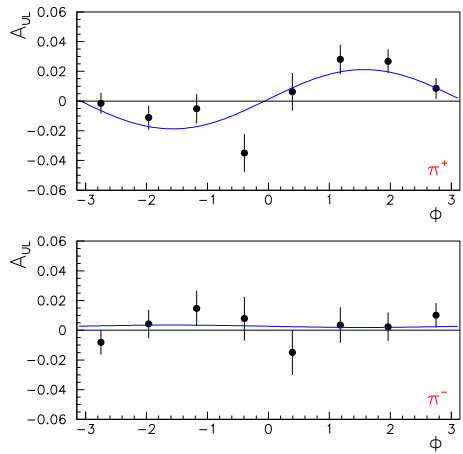,width=6cm}}
\vspace*{8pt}
\caption{Single spin asymmetry $A_{UL}$ for pions as a function of the
Collins angle $\phi$. The curve for $\pi^{+}$ corresponds to the fit
$A=P_{1}+P_{2}sin({\phi}_{\pi})$ with $P_{1}=-0.007\pm 0.003$ and 
$P_{2}=0.020\pm 0.004$. The corresponding curve for $\pi^-$ is given
by $P_{1}=0.003\pm 0.004$ and $P_{2}=-0.001\pm 0.005$. }
\label{azim2}
\end{figure}

HERMES has made the first measurements of single-spin azimuthal asymmetries
for semi-inclusive pion production in DIS, using both unpolarized and
longitudinally polarized proton targets in the HERA $27.5$ GeV
polarized positron storage ring. The kinematic cuts for these measurements
were $1$ GeV$^{2}$ $< Q^{2} < 15$ GeV$^{2}$, $W > 2$ GeV, $0.023 < x <0.4$ 
and $y < 0.85$. Pions
were identified over the range $4.5$ $GeV < E_{\pi} < 13.5$ GeV. Exclusive 
production was suppressed with the requirement $0.2< z <0.7$. The limit
$p_\perp > 50$ MeV was applied to the pions to insure accurate measurement
of the angle $\phi$. Measurements were made with all combinations of
beam and target helicities to permit measurement of single and double
spin asymmetries in the cross section. The kinematics of pion 
electroproduction are presented in Fig.~(\ref{azim1}). Here $k$ and $k'$
are the four momenta of the incoming and outgoing lepton, respectively. 
The transverse momentum ($p_\perp$) of the pion is defined with respect
to the virtual photon direction in the initial photon-proton center-of-mass
system. The angle $\phi$ is the Collins angle which provides the chiral
odd azimuthal dependence in the fragmentation process. The HERMES 
measurements of the single spin asymmetry
\begin{equation}
A_{UL}(\phi)=\frac{1}{P}\frac{N^{+}(\phi)-N^{-}(\phi)}
{N^{+}(\phi)+N^{-}(\phi)},
\label{ssa1}
\end{equation}
where the subscript $UL$ refers to unpolarized beam and longitudinally 
polarized target, provide a clear signature of the Collins effect. The data
are shown in Fig.~(\ref{azim2}).
For $\pi^+$ the distribution shows a strong sin($\phi$) dependence, the 
signature of the Collins effect, while the curve for $\pi ^{-}$ is
isotropic. The difference between the $\pi^{+}$ and $\pi^{-}$ data can
be ascribed to the dominance of $u$ quark fragmentation from the proton to 
the $\pi^{+}$.

Analysis of the data was performed under the assumption that the cross
section factors into a hard scattering term times a fragmentation
function. When this cross section is integrated over the final transverse
momentum of the hadrons, the T-odd terms leading to single-spin asymmetries
vanish. The various contributions to the $\phi$ dependent asymmetry are
isolated by extracting moments of the cross section weighted by the 
corresponding $\phi$ dependent functions. In the present case, sin($\phi$)
was the weighing function of interest. Kotzinian and Mulders\cite{kotz} 
have established that even with a longitudinally polarized target, one 
expects a significant single spin asymmetry with a sin($\phi$) moment
of the form
\begin{equation} 
\langle{sin(\phi)}\rangle _{OL}\propto S_{T}\Sigma_{a}
\, e^{2}_{a}{h^{a}_{1}(x)}{H_{1}^{\perp a}}
+ ...
\label{lead}
\end{equation} 
which arises from the small target spin component transverse to
the direction of the virtual photon. Here, $H_{1}^{\perp a}$ is the Collins 
spin dependent fragmentation function. The analyzing powers for the beam
and target polarizations, $A_{UL}^{sin \phi}$ and $A_{LU}^{sin \phi}$, 
were evaluated by calculating the azimuthially weighted moments of the
event spectra. A substantial analyzing power is observed for $\pi^+$,
$A_{UL}^{sin \phi}=0.022\pm 0.005(stat.)\pm 0.003(syst.)$, while for $\pi^-$,
$A_{UL}^{sin \phi}=0.002\pm 0.006(stat.)\pm 0.003(syst.)$. For both
$\pi^+$ and $\pi^-$ the beam-related analyzing powers $A_{LU}^{sin \phi}$
are consistent with zero, as is the other target related analyzing power
$A_{UL}^{sin 2\phi}$. This is already a strong signal that pion single spin
asymmetries provide a large analyzing power for measuring the transverse
distribution function, $h_{1}(x,Q^{2})$.
\begin{figure}
\centerline{\psfig{file=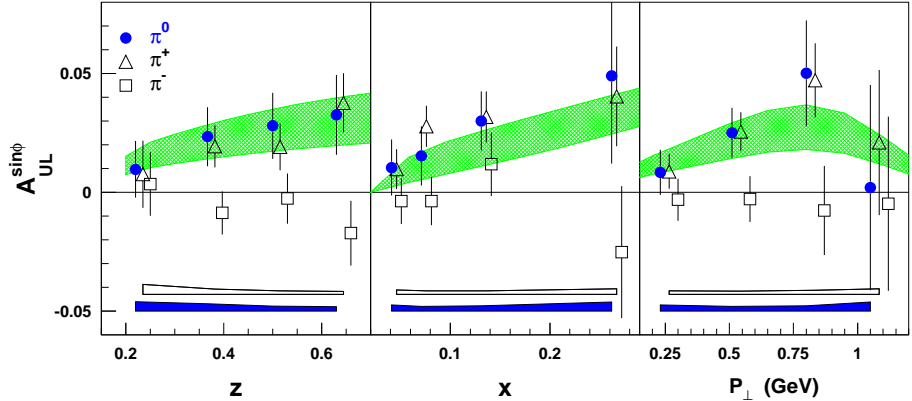,width=12cm}}
\vspace*{8pt}
\caption{Analyzing power $A_{UL}^{sin\phi}$ for pions as a function of 
the fractional pion energy $z$, the Bjorken variable $x$, and the 
pion transverse momentum $p_\perp$. Error bars include only statistical
uncertainties. The open and filled bands at the bottom of the panels
represent the systematic uncertainties for neutral and charged pions,
respectively. The shaded areas show a range of model predictions, 
(see text).}
\label{aul}
\end{figure}

The measurements have been extended to include neutral pions. The results 
are summarized in Fig.~(\ref{aul}) where $A_{UL}^{sin \phi}$ is shown for each
charge as a function of $z$, $x$, and the pion transverse momentum, $p_{\perp}
$, after averaging over the other two kinematic variables. The $\pi^0$ and
$\pi^+$ analyzing powers exhibit similar behaviour in each variable. The 
increase of $A_{UL}^{sin \phi}$ with increasing $x$ suggests that single-spin
asymmetries are valence quark effects. The increase of $A_{UL}^{sin \phi}$
with $p_\perp$ can be related to  the dominant role of intrinsic quark 
transverse momentum when $p_\perp$ remains below about 1 GeV/c.  
The results for $\pi^+$ and $\pi^0$ follow the predictions of a model
calculation\cite{sanc} in which the 
distribution function of Eq.~(\ref{lead}) is 
approximated by $h_1$. The range of predictions shown in Fig.~(\ref{aul})
correspond to the limits $h_{1}=g_{1}$ (non-relativistic limit) and
$h_{1}=(f_{1}+g_{1})/2$ (Soffer inequality), where $g_{1}$ and $f_{1}$ 
are the usual polarized and unpolarized distribution functions. The 
T-odd fragmentation function $H_{1}^{\perp (1)}(z)$ was assumed to follow
the Collins parameterization\cite{coll}. The success of this simple model 
calculation provides confidence that accurate determinations of transversity
distributions are possible through observation of pion single spin
asymmetries with transversely polarized targets. Such measurements are
scheduled by the HERMES collaboration in the immediate future.

\section{Deep Virtual Compton Scattering}

\begin{figure}[b]
\centerline{\psfig{file=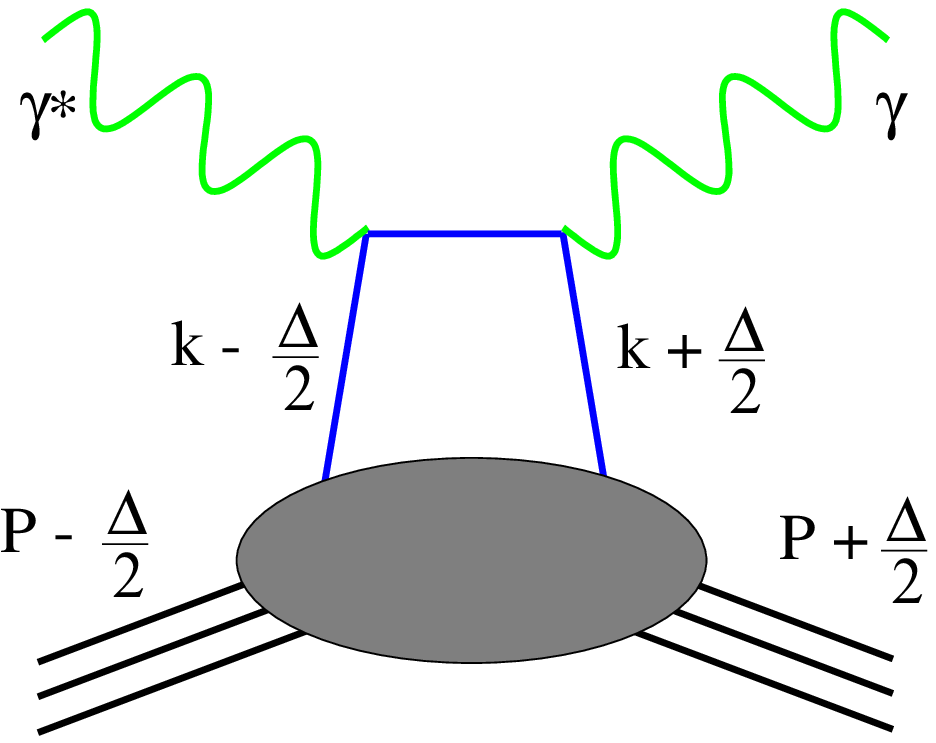,width=4cm}\hspace*{2cm}
\psfig{file=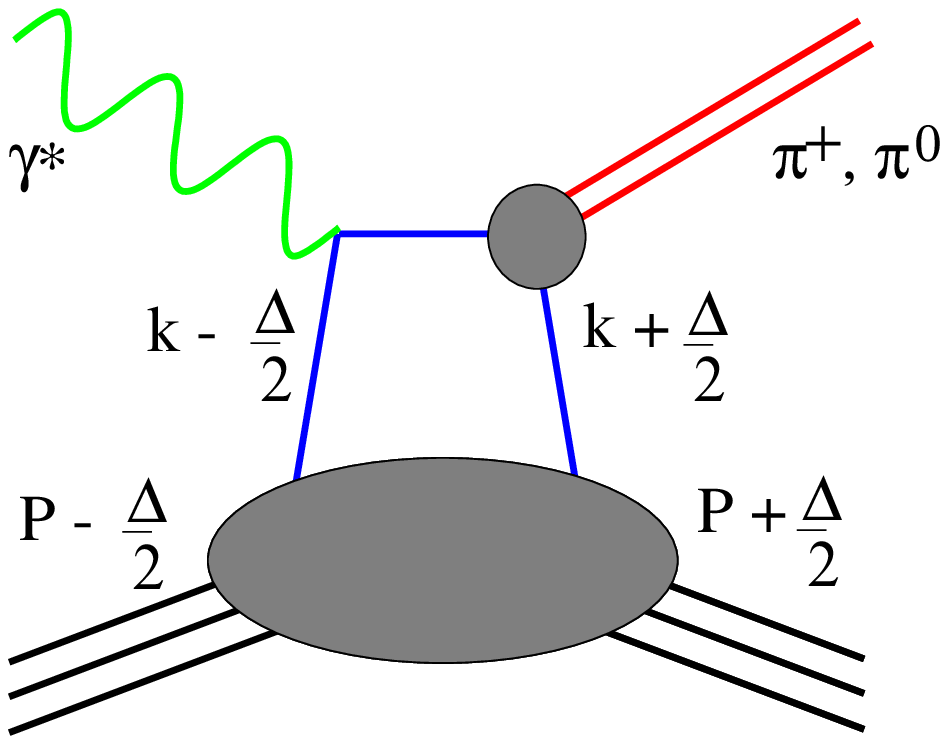,width=4cm}}
\vspace*{8pt}
\caption{Graphic representation of processes involving
generalized parton distributions,
deep virtual Compton scattering (left) and exclusive pion production (right).
The GPD describes the emission of a parton of momentum $k-\frac{\Delta}{2}$
and its reabsorption with momentum $k+\frac{\Delta}{2}$.}
\label{gpd1}
\end{figure}
Deep virtual Compton scattering (DVCS) can be viewed as the scattering
of a virtual photon generated in lepton scattering into the continuum.
Although little data on the process are available, the process is of wide 
interest because exclusive DVCS is the simpliest example of a reaction 
described by a new class of parton
distributions, i.e. generalizations of the usual forward parton 
distributions (pdf's). While ordinary pdf's give the probability of
finding a quark with a momentum fraction $x=k/p$ 
in the nucleon,
generalized parton distributions (GPD's) describe the removal of a 
quark $q(k-\frac{\Delta}{2})$ and implantation of $q'(k+\frac{\Delta}{2})$.
The example of DVCS is shown graphically in Fig.~(\ref{gpd1}). Factorization
theorms have been formulated\cite{coll2} 
which demonstrate that contributions for
hard exclusive reactions have the form of products of GPD's times hard 
scattering coefficients which are calculable from perturbative QCD.   
There are four families of GPD's, one pair of unpolarized distribution 
functions, $H^{q}(x,\xi ,t)$ and 
$E^{q}(x,\xi ,t)$, and a second pair of polarized
distribution functions, $\tilde{H}^{q}(x,\xi ,t)$ 
and $\tilde{E}^{q}(x,\xi ,t)$. 
The relevant light cone variables are the longitudinal momentum fraction 
$x$, the skewness variable $\xi =-\Delta^{+}/2P^{+}$, and the four momentum
transfer to the nucleon squared $t=\Delta^{2}$. $H^{q}$ and $\tilde{H}^{q}$
conserve the helicity of the nucleon, while $E^{q}$ and $\tilde{E}^{q}$ 
describe processes in which the nucleon helicity may flip. GPD's combine 
the character of ordinary parton distributions and nucleon form factors 
through their kinematic limits and moments, 
e.g. $q(x)=H^{q}(x,\xi =0,t=0)$, and 
$\Delta q(x)=\tilde{H}^{q}(x,\xi =0,t=0)$. The first moments constrain 
nucleon form factors through the relations $\int_{-1}^{1}dxH^{q}(x,\xi ,
\Delta ^{2})=F_{1}^{q}(\Delta ^{2} )$,
$\int_{-1}^{1}dxE^{q}(x,\xi ,\Delta ^{2})=F_{2}^{q}(\Delta ^{2})$,
$\int_{-1}^{1}dx\tilde{H}^{q}(x,\xi ,\Delta ^{2})=g_{A}^{q}(\Delta ^{2} )$,
and $\int_{-1}^{1}dx\tilde{E}^{q}(x,\xi ,\Delta ^{2})=h_{A}^{q}(\Delta ^{2})$.
$F_{1}(t)$ and $F_{2}(t)$ are the Dirac and Pauli form factors, while 
$g_{A}(t)$ and $h_{A}(t)$ are the axial vector and pseudoscalar form factors,
respectively. The strong interest in GPD's comes, in part, from the
observation\cite{Ji} that their second moments can be connected  to the
total orbital angular momentum carried by the partons through the relation
\begin{equation}
J^{q}=\frac{1}{2}\Delta\Sigma + L^{q}=\frac{1}{2}\int^{1}_{-1}
xdx[H^{q}(x,\xi ,t=0)+E^{q}(x,\xi,t=0)].
\label{jtot}
\end{equation}  
Thus, GPD's may provide a measure of the $J^{q}$ and therefore with prior
knowledge of $\Delta\Sigma$, of the 
orbital angular momentum of the partons, $L^{q}$.

DVCS is coherent with Bethe-Heitler bremstrahlung, and the interference 
between these two processes provides the opportunity to isolate DVCS
amplitudes. The cross section for exclusive photon production is of the 
form\cite{beli}
\begin{equation}
\frac{d^{4}\sigma}{d\phi dtdQ^{2}dx}=\frac{xy^2}{32(2\pi )^{4}Q^{4}}
\frac{|\tau_{BH}+\tau_{DVCS}|^2}{(1+4x^{2}m^{2}/Q^{2})^{1/2}}.
\label{sigdv}
\end{equation} 
Both the magnitude and phase of the interference term can be measured with
an unpolarized target. Measurements of the charge asymmetry with an
unpolarized beam give
\begin{equation}
d\Delta\sigma_{ch}\equiv d\sigma_{e^+}-d\sigma_{e^-}\sim cos({\phi})
\times Re(\tau_{DVCS}\tau_{BH}),
\label{chas}
\end{equation}
while measurements of the spin asymmetry with respect to the beam helicity
give
\begin{equation}
d\Delta\sigma_{LU}\equiv d\sigma_{\leftarrow}-d\sigma_{\rightarrow}
\sim sin({\phi})\times Im(\tau_{DVCS}\tau_{BH}).
\label{beas}
\end{equation}
Thus, these asymmetries access the real and imaginary parts of the same
interference amplitude which has the form 
\begin{equation}
\tau_{DVCS}\tau_{BH}\propto F_{1}\mathcal{H}_{1}+\frac{x_B}{2-x_B}
(F_{1}+F_{2})\tilde{\mathcal{H}}_{1}-\frac{\Delta_2}{4M^2}
F_{2}\mathcal{E}_{1}.
\label{inta}
\end{equation}
The DVCS amplitudes of Eq.~(\ref{inta}), 
$\mathcal{H}_{1}$,$\tilde{\mathcal{H}}_{1}$, and $\mathcal{E}_1$,
can be given as convolutions\cite
{beli} in $t$ of perturbative calculable hard scattering parts with GPD's
for DVCS.

\begin{figure}[t]
\centerline{\psfig{file=figure2-dvcs.eps,width=6.5cm}
\psfig{file=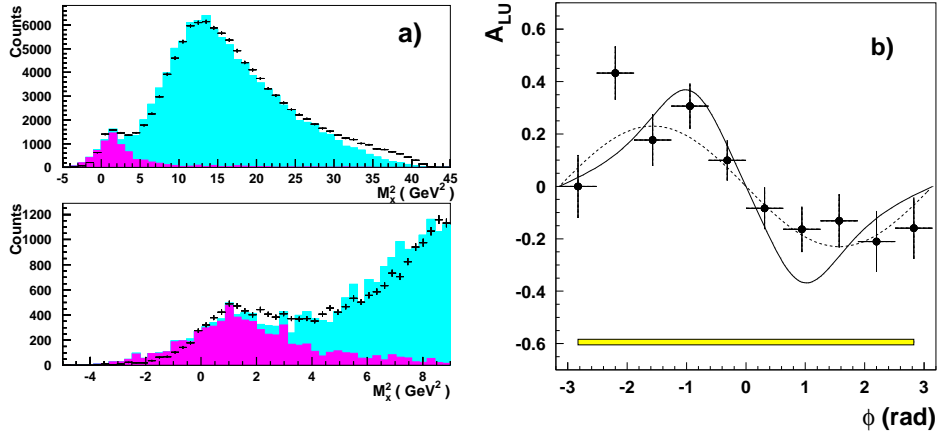,width=6cm}}
\vspace*{8pt}
\caption{a) The measured distribution of photons observed in hard
electroproduction as a function of the missing mass squared $M_{x}^2$.
The upper panel shows the full kinematic range, while the low $M_{x}^2$ 
domain is presented in the lower panel. The light-gray histogram 
represents the results of a Monte Carlo simulation 
of the experiment which includes 
fragmentation processes and the Bethe-Heitler process, while the dark-shaded
histogram represents only the Bethe-Heitler contribution.
b) Beam-helicity analyzing power $A_{LU}$ as a function of the azimuthal angle
$\phi$. The dashed curve represents a sin$\phi$ dependence with an
amplitude of $-0.23$, while the solid curve is a theoretical calculation in
the approximation of small $t$ (see text).}
\label{mmsp}
\end{figure}
In HERMES DVCS has been measured with the HERA longitudinally
polarized positron beam at DESY, with longitudinally polarized and
unpolarized targets. Events were selected if they contained only one 
positron with a momentum larger than $3.5$ GeV, and only one photon
with an energy greater that $0.8$ GeV in the calorimeter. Kinematic 
cuts of $Q^{2}>1$ GeV$^2$, $W^{2}>4$ GeV$^2$, and $\nu < 24$ GeV were 
imposed. In addition,  the cut $15<\Theta_{\gamma *\gamma}<70$ mrad
was used to avoid false asymmetries from bias in reconstruction of small
angles. A contamination of events from $\pi^{0}\rightarrow 2\gamma$ decay
which arises from the calorimeter granularity was determined to be
$<5\%$. The missing mass spectrum for DVCS is shown in Fig.~(\ref{mmsp}), 
where it is compared to results of a Monte Carlo simulation which 
includes photons generated from DIS as well as those resulting from the
exclusive Bethe-Heitler process, $e+p\rightarrow e'+p+\gamma$. Because of
the finite resolution of the spectrometer, $M_{x}^2$ may be negative, in
which case by definition, $M_{x}=-\sqrt{-M_{x}^2}$. The missing mass region
between $-1.5$ and $+1.7 GeV$ was selected for isolating exclusive DVCS. 
This region contains both exclusive scattering to the proton ground state
and the $\Delta (1232)$ resonance. However, because the Bethe-Heitler process
is strongly dominated by the elastic channel the spin asymmetries are 
expected to have a small contribution from the $\Delta (1232)$ resonance.

The azimuthal distribution of events in the missing mass window
centered on the proton mass is presented in Fig.~(\ref{mmsp}) where the 
beam-helicity analyzing power
\begin{equation}
A_{LU}(\Theta )=\frac{1}{\langle|P_{b}|\rangle}\frac{N^{+}(\phi)-N^{-}(\phi )}
{N^{+}(\phi)+N^{-}(\phi )}
\label{anpw}
\end{equation}
is plotted. The data show the characteristic sin$\phi$ dependence expected
from Eq.~(\ref{beas}). The solid curve of Fig.~(\ref{mmsp}) is the result
of a calculation\cite{kive} with GPD's in the approximation of small $t$
where factorization into contributions from form factors and parton 
distributions is reasonable. The calculation gives a good description of
the data.
\begin{figure}
\centerline{\psfig{file=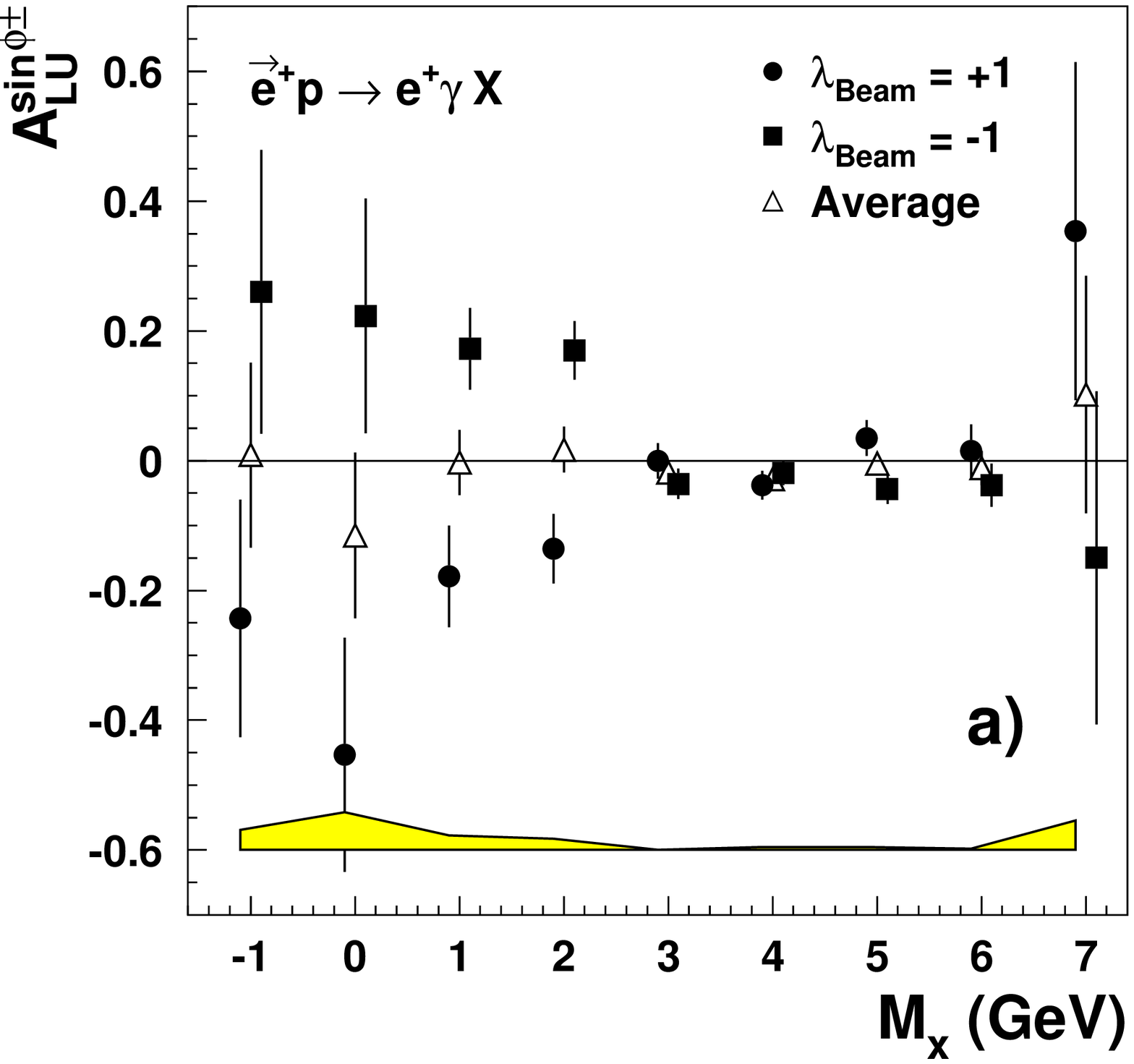,width=6cm}
\psfig{file=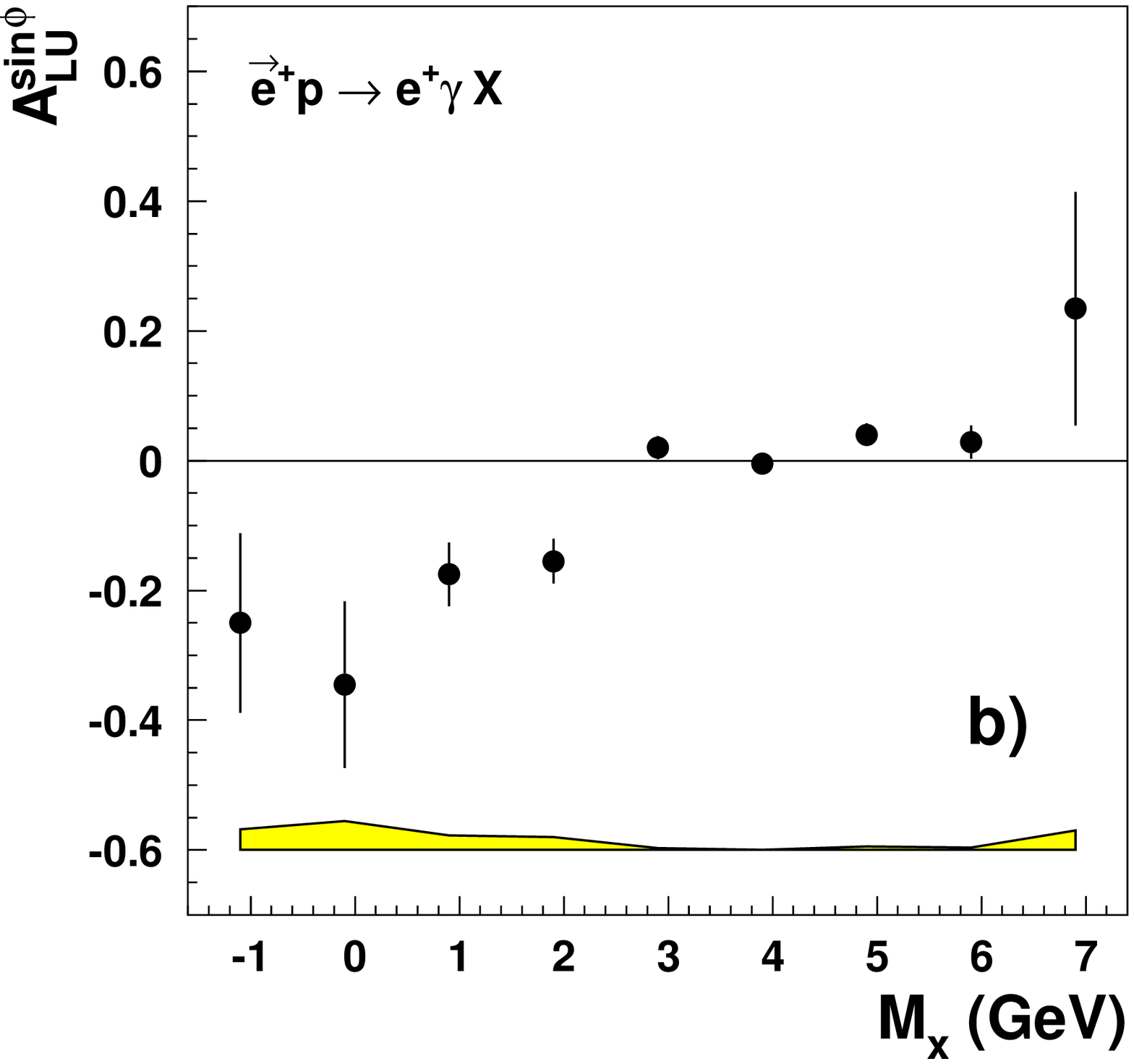,width=6cm}}
\vspace*{8pt}
\caption{a) The sin$\phi$-moment $A_{LU}^{sin\phi^\pm}$ as a function
of the missing mass for positive beam helicity (circles), negative
beam-helicity (squares) and the averaged helicity (open triangles).
b) The beam-helicity analyzing power $A_{LU}^{sin\phi}$ as a function
of the missing mass.}
\label{ssas}
\end{figure}
Checks on the beam-helicity and missing mass dependence were performed by
examining the distributions of the sin$\phi$-weighted moments which were
defined as
\begin{equation} 
A^{sin{\phi}^{\pm}}_{LU}=\frac{2}{N^\pm}\sum^{N^{\pm}}_{i=1}\frac
{sin\phi _{i}}{|P_{b}|_{i}}.
\label{mom1}
\end{equation}
The sin$\phi$ moments are shown in Fig.~(\ref{ssas}) for the two beam 
helicities together with their average. As expected, the sign of the moment
is opposite for the two helicities and their average is consistent with zero.
Also, as expected the moments vanish for high missing mass. The data for 
the two beam helicities can be combined to determine the beam-helicity
analyzing power
\begin{equation}
A^{sin{\phi}}_{LU}=\frac{2}{N}\sum^{N}_{i=1}\frac
{sin\phi _{i}}{(P_{b})_{i}}
\label{mom2}
\end{equation}
Combining the data in the $M_x$ region between $-1.5$ and $+1.7$ GeV 
yields the result $A^{sin\phi}_{LU}=
-0.23\pm 0.04(stat.)\pm 0.03(system)$ which is to be compared with the
theoretical calculation\cite{kive} of $A^{sin\phi}_{LU}=-0.37$. The
average values of the kinematic variables corresponding to this 
measurement are $<x>=0.11$, $<Q^{2}>=2.6$ GeV$^2$, and $<t>=0.27$ GeV$^2$. 
The HERMES results provide a clear demonstration of a strong experimental
signature for DVCS which can be exploited to study GPD's in what
may be the conceptually clearest case. The measurement described here 
is the initial step in a program of systematic studies, but the results
conform to theoretical expectations and foretell a new rich area of study.

\section{The Future}

In spite of many years of experiments, nucleon spin structure
remains a complex and subtle problem. A detailed decomposition of the
spin of the nucleon remains elusive. However, as the data from the HERMES
experiment indicate, we are beginning to obtain information on some
of the central questions. What is the polarization of the gluons and the 
strange sea? How are transverse and longitudinal spin distributions 
related? Future measurements at HERMES will focus on direct measurements of 
transversity with transversely polarized targets. Measurements planned
in the RHIC spin program\cite{sait,vigd} 
will provide the first detailed probing of 
gluon spin distributions. Measurements of DVCS may provide 
sufficient access to GPD's to enable mapping out their variation with kinematic
variables $(x,\xi ,t)$, posing a major challenge to experimenters.
Will accurate determination of the second moments of the GPD's be possible?
Can we access the parton angular momenta $L_q$ with measurements of
DVCS?

\section*{Acknowledgements}

The support of the DESY management and staff and the staffs of the
collaborating institutions is gratefully acknowledged. The author wishes
to thank R. J. Holt, G. van der Steenhoven, and P. Reimer for a careful
reading of the manuscript. The author also acknowledges the massive 
efforts of all the HERMES collaborators which have made the program
a success. This work was supported in part by  
the U.S. Department of Energy and National Science Foundation.

\end{document}